\documentclass{article}
\usepackage{arxiv}
\usepackage[numbers,sort&compress]{natbib}
\usepackage[utf8]{inputenc} 
\usepackage[T1]{fontenc}    
\usepackage{hyperref}       
\usepackage{url}
\usepackage{textcomp}
\usepackage{ gensymb }
\usepackage[ruled,vlined]{algorithm2e}
\usepackage{algorithmic}
\usepackage{booktabs}       
\usepackage{amsfonts}       
\usepackage{nicefrac}       
\usepackage{microtype}      
\usepackage{lipsum}
\usepackage{float}
\usepackage{graphicx}
\usepackage{multirow}
\usepackage{array}
\usepackage{comment}
\usepackage{longtable}
\usepackage{ragged2e}
\usepackage{subfig}

\title{Monkeypox Image Data collection}

\author{
  Md Manjurul Ahsan \\
  Industrial and Systems Engineering\\
  University of Oklahoma\\
  Norman, Oklahoma-73071 \\
  \texttt{ahsan@ou.edu} \\
   \And
 Muhammad Ramiz Uddin \\
  Dept. of Chemistry and Biochemistry\\
  University of Oklahoma\\
  Norman, Oklahoma-73019\\
  \texttt{muhammadramizuddin@gmail.com}\\
  \And
 Shahana Akter Luna \\
  Medicine \& Surgery\\
  Dhaka Medical College \& Hospital\\
  Dhaka, Bangladesh- 1000\\
  \texttt{shahanaakterluna123@gmail.com}\\} 

\begin{document}
\maketitle

\begin{abstract}
This paper explains the initial Monkeypox Open image data collection procedure. It was created by assembling images collected from websites, newspapers, and online portals and currently contains around 1905 images after data augmentation.
\end{abstract}


\section{Motivation}
In the context of the recent crisis led by Monkeypox~\cite{dye2022investigating}, it is critical to develop and test machine learning-driven diagnosis models due to the unavailability of the datasets. While there exist large datasets for various skin diseases such as chickenpox, rash, etc.~\cite{Kaggle2022}, there is no collection of  Monkeypox disease-related image datasets designed to be used for computational analysis. Our data collection approach is highly motivated by Dr. Joseph Cohenes's initial small dataset creation during the onset of COVID-19, which contains 123 image samples at the early stage when the datasets were publicly available~\cite{cohen2020covid}.

This paper introduces a public dataset of Monkeypox disease cases with Chickenpox, Meares which infected the hand, face, leg, and neck of the human body (mostly infected due to the virus).
Our collaborative team believes that our dataset will dramatically initiate the initial image-based diagnosis research. Notably, this would provide essential data to train and test deep learning models such as convolutional neural networks, generative adversarial networks, and transfer learning approaches.

Currently, all images are available on the following GitHub repository link:\url{https://github.com/mahsan2/Monkeypox-dataset-2022}.
\section{Expected outcomes}
This dataset can be used to study the progress of Monkeypox and how its image-based findings vary from other types of skin diseases. We expect our small dataset will play a significant role in developing the Monkeypox diagnosis tool and its predicted outcomes.

Tools can be built and used to limit the dependency on traditional microscopic image analysis and polymerase chain reactions (PCR)~\cite{ahsan2020covid,ahsan2021detecting} to identify Monkeypox diseases~\cite{ahsan2020deep}. This tool might reduce the contact between the healthcare practitioners and the patients. Further, it is also possible to develop mobile-based disease diagnosis systems that the patient can use~\cite{sallam2019mobile}.
\section{Dataset}
\begin{figure}
    \centering
    \includegraphics[width=\textwidth]{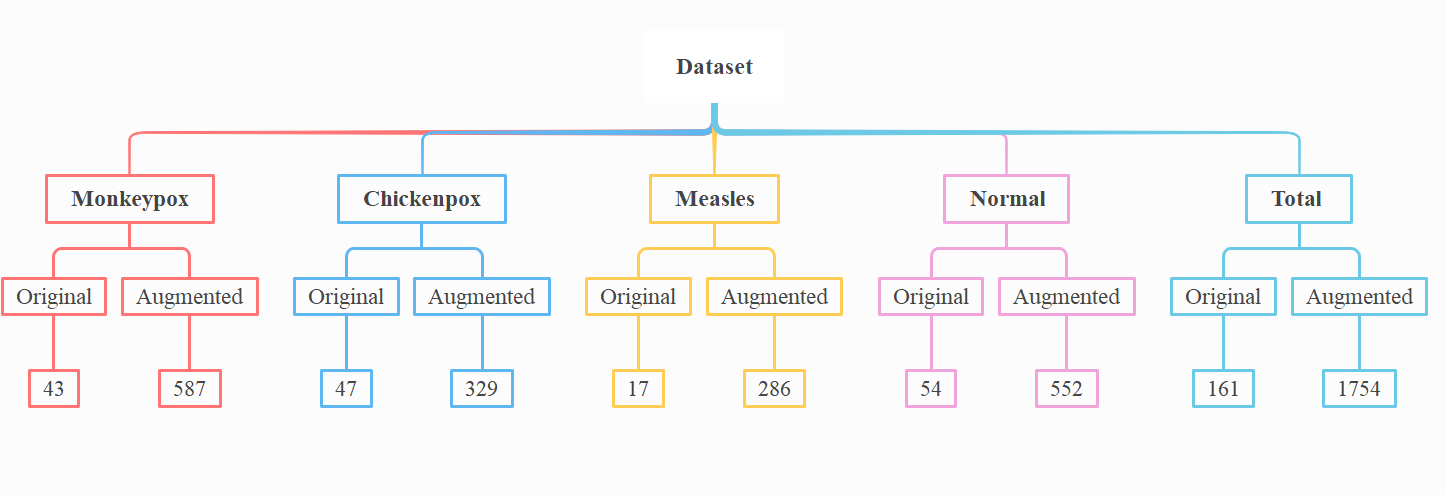}
    \caption{Characteristics of the current datasets after dataset development.}
    \label{fig:dataset}
\end{figure}

Figure~\ref{fig:dataset} shows the characteristics of the current datasets. The dataset contains original 43 Monkeypox, 47 Chickenpox, 17 Measles, and 54 Normal image samples of individuals. Further additional data is added using data augmentation techniques within the scope of the facility provided by the Keras Image data generator~\cite{tensorflow2020,sreene2020}.

\begin{figure}[ht]
    \centering
    \includegraphics[width=\textwidth]{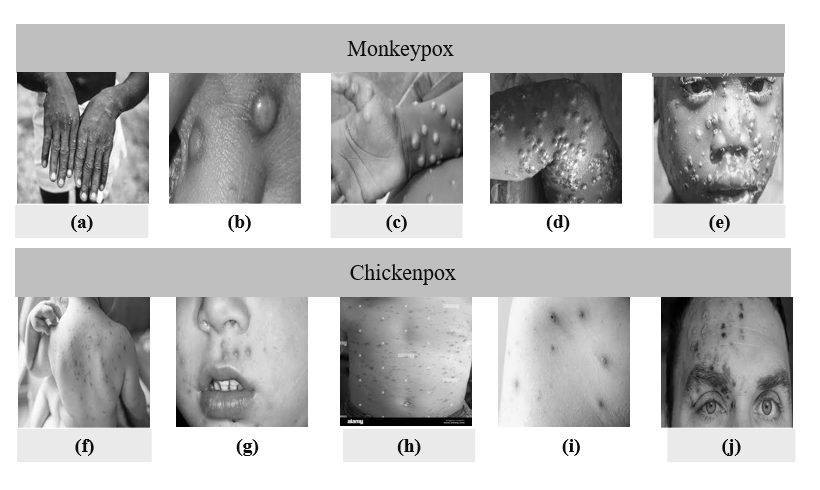}
    \caption{Demonstration of some of the images from the datasets.}
    \label{fig:dataf}
\end{figure}

Figure~\ref{fig:dataf} displays some of the sample images from the newly created datasets. Figure~\ref{fig:dataf}(a)-(e) contains Monkeypox disease patient's images, wherein Figure~\ref{fig:dataf}(f)-(j) contains images of Chickenpox disease patients.

\begin{figure}
    \centering
    \includegraphics[width=\textwidth]{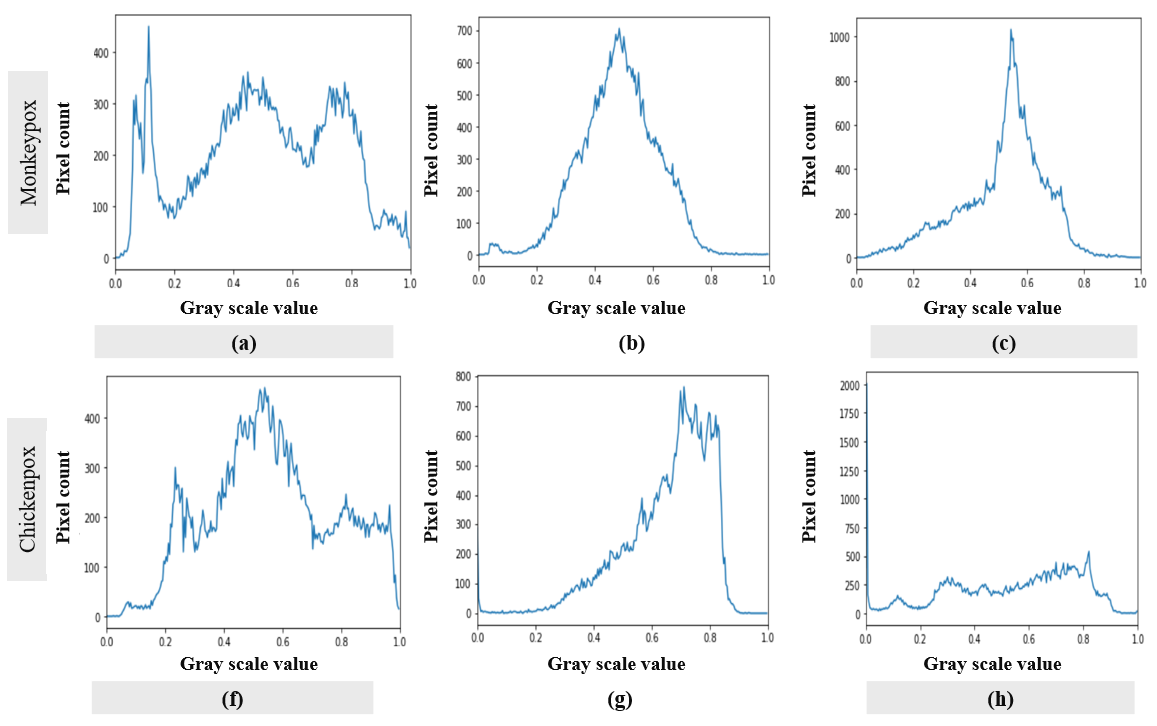}
    \caption{Histogram of image data available within the datasets.}
    \label{fig:histo}
\end{figure}

Histogram analysis is one of the best ways to identify the pixel intensity and differences among various images. Figure~\ref{fig:histo} plots the histogram for some of the samples illustrated in Figure~\ref{fig:dataf}. Figure~\ref{fig:histo}(a)-(c) plotted a histogram of the corresponding images of Figure~\ref{fig:dataf}(a)-(c), wherein Figure~\ref{fig:histo}(f)-(h) plotted the subsequent histogram of Figure~\ref{fig:dataf}(f)-(h).
\bibliographystyle{unsrt}  
\bibliography{main}

\end{document}